\documentclass[pra,twocolumn,showpacs,superscriptaddress,amsmath]{revtex4}
\usepackage{graphicx}

\newcommand{\ket}[1]{\, | #1 \rangle}
\newcommand{\braket}[2]{\langle #1 | #2 \rangle}
\newcommand{\expv}[1]{\langle #1 \rangle}
\newcommand{\ga}{\gamma}

\newcommand{\de}{\delta}

\newcommand{\ve}{\varepsilon}
\newcommand{\us}{\uparrow}
\newcommand{\ds}{\downarrow}
\newcommand{\bk}{b^{\dagger}}
\newcommand{\ck}{c^{\dagger}}
\newcommand{\be}{\begin{equation}}
\newcommand{\ee}{\end{equation}}
\newcommand{\bea}{\begin{eqnarray}}
\newcommand{\eea}{\end{eqnarray}}
\newcommand{\besa}{\begin{subequations}\begin{eqnarray}}
\newcommand{\eesa}{\end{eqnarray}\end{subequations}}

\begin{document}


\title{Quantum liquid of repulsively bound pairs of particles in a lattice}

\author{David Petrosyan} 
\affiliation{Institute of Electronic Structure \& Laser, FORTH, 
71110 Heraklion, Crete, Greece}
\author{Bernd Schmidt}
\affiliation{Fachbereich Physik, Technische Universit\"at Kaiserslautern, 
D-67663 Kaiserslautern, Germany}
\author{James R. Anglin}
\affiliation{Fachbereich Physik, Technische Universit\"at Kaiserslautern, 
D-67663 Kaiserslautern, Germany}
\author{Michael Fleischhauer}
\affiliation{Fachbereich Physik, Technische Universit\"at Kaiserslautern, 
D-67663 Kaiserslautern, Germany}

\date{\today}

\begin{abstract}
Repulsively interacting particles in a periodic potential can 
form bound composite objects, whose dissociation is suppressed
by a band gap. Nearly pure samples of such repulsively bound 
pairs of cold atoms---``dimers''---have recently been prepared
by Winkler {\it et al.} [Nature {\bf 441}, 853 (2006)]. We here 
derive an effective Hamiltonian for a lattice loaded with dimers 
only and discuss its implications to the many-body dynamics of the
system. We find that the dimer-dimer interaction includes strong
on-site repulsion and nearest-neighbor attraction which always 
dominates over the dimer kinetic energy at low temperatures. 
The dimers then form incompressible, minimal-surface ``droplets''
of a quantum lattice liquid. For low lattice filling, the effective
Hamiltonian can be mapped onto the spin-$\frac{1}{2}$ $XXZ$ model
with fixed total magnetization which exhibits a first-order phase 
transition from the ``droplet'' to a ``gas'' phase. This opens the 
door to studying first order phase transitions using highly 
controllable ultracold atoms. 
\end{abstract}


\maketitle

\section{Introduction}

Many-body systems in spatially periodic potentials have been 
investigated since the early days of quantum theory \cite{SolStPh}. 
Idealized descriptions such as the Heisenberg spin and Hubbard models 
have been used to examine basic issues in condensed matter physics. 
The Hubbard model has recently acquired direct experimental 
significance, however, describing cold atomic gases trapped in 
optical lattices \cite{OptLatRev}. The relevant parameters of 
these systems can be tuned to implement the Hubbard model with a 
remarkable accuracy. In particular, the transition from the superfluid
to the Mott insulator phase \cite{bosMI} with a commensurate number of 
bosonic atoms per site has been demonstrated \cite{optlattMIth,optlattMIex}.
More recently, Winkler {\it et al.} \cite{KWEtALPZ} have observed another 
lattice effect: binding of {\it repulsively} interacting bosons into close
pairs which are dynamically stable in the absence of dissipation. 
Repulsively bound composite objects are a general phenomenon, appearing 
in various periodic systems possessing a band gap at the relevant 
``dissociation'' energy. Electrons have been shown to pair, via Coulomb 
repulsion, in arrays of tunnel-coupled quantum dots \cite{GNDPPL}. 
Analogous effects have been predicted for strongly interacting mixtures 
of bosonic and fermionic atoms in an optical lattice  \cite{LSBF}, or photons
forming gap solitons in nonlinear photonic bandgap structures \cite{Gershon}.

Here we study a lattice loaded with even numbers of bosonic atoms at each 
site, in the experimentally relevant regime \cite{KWEtALPZ} where the on-site 
repulsion between atoms exceeds the inter-site tunneling rate. We derive 
an effective Hamiltonian for repulsively bound atom pairs (``dimers''), 
which describes the many-body dynamics of the pairs, viewed as single 
composite objects. A special property of the system, not shared by 
less exotic systems, is the effective occupation-dependent tunneling and 
nearest-neighbor interactions of the dimers mediated by the single-atom 
tunneling via nonresonant virtual states. We find that the attractive 
interaction between the dimers always exceeds their kinetic energy which,
combined with the still stronger on-site repulsion that the dimers 
inherit from the repulsion among their constituent atoms, leads to 
clustering of the dimers into ``droplets'' with minimum surface area 
and uniform density. When the system is initially prepared with at 
most one dimer per site \cite{KWEtALPZ}, the effective Hamiltonian 
can be mapped onto the spin-$\frac{1}{2}$ $XXZ$ model \cite{Takahashi}
with fixed magnetization, which is known to exhibit a first order phase
transition from a ``droplet'' to a ``gas'' phase  at a critical temperature
\cite{Sadchev}.

\section{Repulsively bound dimer within the Bose-Hubbard model}

The dynamics of cold bosonic particles occupying the lowest Bloch band 
of a tight-binding periodic potential is governed by the Bose--Hubbard
Hamiltonian ($\hbar = 1$)
\be
H = \sum_{j} \ve_{j} \hat{n}_j + 
\frac{U}{2} \sum_{j} \hat{n}_j (\hat{n}_j -1) 
- J \sum_{\expv{j,i}} \bk_{j} b_{i} , \label{HubHam}
\ee
where $\bk_{j}$ ($b_{j}$) is the creation (annihilation) 
operator for a boson at site $j$ with energy $\ve_{j}$, 
$U$ is the on-site interaction (repulsion for $U > 0$),  
$\hat{n}_j = \bk_{j} b_{j}$ is the number operator for site $j$,
and $J$ the tunneling rate between adjacent sites $\expv{j,i}$. 
A natural basis for the Hamiltonian (\ref{HubHam}) is that of the 
eigenstates $\ket{n_j} \equiv \frac{1}{\sqrt{n !}} (\bk_j)^{n} \ket{0}$
of the number operator $\hat{n}_j$ whose eigenvalues $n = 0,1,2,\ldots$ 
denote the number of particles at site $j$, and 
$\ket{0} \equiv \ket{\{ 0_j \}}$ is the vacuum state. 
For a single particle in a uniform ($\ve_{j} = \ve$ for all $j$)
periodic potential, the on-site interaction plays no role, and resonant 
tunneling leads to a  Bloch band of width  $4 d J$ centered around $\ve$, 
where $d$ is the system dimension. 

Considering next two particles in a periodic potential, according 
to Eq. (\ref{HubHam}), the state $\ket{2_j}$ with two particles 
localized at the same site has an energy offset $U$ from the state 
$\ket{1_j}\ket{1_i}$ with $i \neq j$. The transition between states 
$\ket{1_j}\ket{1_i}$ and $\ket{2_j}$ is therefore non-resonant and 
is suppressed when $U \gg J$. If initially the particles occupy 
different sites, each particle can tunnel freely from site to site, 
until it encounters the other particle at a neighboring site. At this
point the two particles undergo elastic scattering and separate again, 
since the maximal kinetic energy $4 d J$ of the two particles is below 
the potential barrier $U$ associated with two particles occupying the 
same site. Note that, in second-order in the small parameter $J/U$,
an adiabatic elimination of the nonresonant states $\ket{2_j}$
and $\ket{2_i}$ yields an effective energy shift of state
$\ket{1_j}\ket{1_i}$ with two particles at the adjacent sites 
$\expv{j,i}$, given by  $-4J^2/U$. This effective attraction 
between a pair of particles at the neighboring sites is, however,
small compared to the single-particle tunneling rate $J$, and 
therefore can not bind the particles together. Conversely, if the 
system is initially prepared in state $\ket{2_j}$, then in order 
for the two particles to separate ($\ket{2_j} \to \ket{1_j}\ket{1_i}$)
via the last term of Eq.~(\ref{HubHam}), energy of the order of $U$ 
would have to be discarded. In the absence of dissipation, this is 
not possible, so the two particles are repulsively bound as a dimer 
\cite{KWEtALPZ}. Using the perturbative analysis outlined in  
the Appendix \ref{app:1dmrwv}, it is easy to show that the localization
(or ``bond'') length of the dimer is $\zeta = [2\ln (U/d J)]^{-1}$, 
so that $\zeta<1$ for $U / J > d \sqrt{e}$. Hence, the dimer constituents
are strongly co-localized for $U / J \gg 1$, which will be assumed from now on.

An important aspect of the problem is the dimer mobility. Although the 
first-order transition $\ket{2_j} \to \ket{1_j}\ket{1_i}$ (with $j$ and 
$i$ denoting adjacent sites) effected by the last term of Eq.~(\ref{HubHam}) 
is nonresonant, in the second order in $J$, the transition 
$\ket{2_j} \to \ket{2_i}$ via the virtual intermediate state 
$\ket{1_j}\ket{1_i}$ is resonant. An adiabatic elimination 
\cite{adelim} of the intermediate state $\ket{1_j}\ket{1_i}$ then 
yields an effective tunneling rate for a dimer as a whole, given by 
$J^{(2)} \equiv 2 J^2 /U \ll J$. Note also that the adiabatic elimination 
of nonresonant state $\ket{1_j}\ket{1_i}$ results in an energy shift
of the dimer state $\ket{2_j}$ equal to $J^{(2)}$, which constitutes
a correction to the dimer energy $2 \ve + U$. Since the dimer is 
surrounded by $2d$ empty sites, each shifting its energy by $J^{(2)}$,
the dimer energy becomes $2 \ve + U + 2 d J^{(2)}$. In analogy with a single 
particle case, the effective tunneling with the rate $J^{(2)}$  
implies a narrow Bloch band for single dimers, of width $4 d J^{(2)}$ 
centered around $2 \ve + U + 2 d J^{(2)}$.

\section{Effective Hamiltonian for a system of dimers}

So far, we have discussed the properties of a single repulsively
bound dimer in a periodic potential. Our aim next will be to describe 
the dynamics of a system of dimers. We know no useful exact analytic 
treatments, but for $J\ll U$ the perturbation approach outlined in the 
Appendix \ref{app:1dmrwv} can be extended to this problem straightforwardly.
Expressing the particle number as $n = 2 m$, where $m = 0,1,2\ldots$ 
represents the number of dimers at a given site, we denote the state
containing $m$ dimers at site $j$ as $\ket{m_j^D}$.
It is convenient to define the operators
$c_j = \frac{1}{\sqrt{2(\hat{n}_j +1)}} \, b_j^2$, and
$\ck_j = (\bk_j)^2 \frac{1}{\sqrt{2(\hat{n}_j +1)}}$,
which annihilate and create a dimer at site $j$. Within the subspace 
of states in which all occupation numbers are even, these operators 
behave exactly as canonical creation and annihilation operators,
possessing the standard bosonic commutation relations
$[c_j , \ck_i] = \de_{ji}$ and $[c_j , c_i] = [\ck_j , \ck_i] = 0$.
The dimer number operator at site $j$ is then given by 
$\hat{m}_j = \ck_j c_j = \hat{n}_j /2$. It is easy to 
verify by induction that 
$\ket{m_j^D} \equiv \frac{1}{\sqrt{m !}} (\ck_j)^{m } \ket{0}$.

\begin{figure}[t]
\centerline{\includegraphics[width=8.7cm]{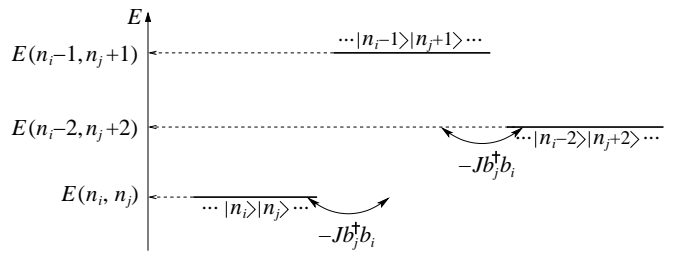}}
\caption{Energy level diagram and tunnel couplings employed in the 
adiabatic elimination of nonresonant states with odd occupation numbers.
$\cdots |n_i\rangle|n_j\rangle \cdots $ denotes a state
with $n_i = 2 m_i$ bosons at site $i$ and $n_j = 2 m_j$ bosons 
at site $j=i+1$ etc., while the energy is
$E(n_i, n_j) = \frac{U}{2} \big[n_i (n_i-1) + n_j (n_j-1) \big] + C$,
where $C$ contains the energy of all the other sites.}
\label{fig:lvldiagr}
\end{figure}

We can now derive an effective Hamiltonian $H_{\mathrm{eff}}$
for a periodic potential loaded with dimers only \cite{KWEtALPZ}.
To zeroth order in the tunneling interaction $J$, only the first two 
terms of Eq.~(\ref{HubHam}) survive. In terms of the dimer operators 
we have defined, they are given, respectively, by $2 \ve \sum_{j} \hat{m}_j$
and $U \sum_{j} \hat{m}_j \big( 2 \hat{m}_j - 1 \big)$. To first order in $J$,
under the condition of the strong on-site repulsion $U \gg J$, all the 
states containing odd numbers of particles per site will be 
non-resonant, and can be eliminated adiabatically \cite{adelim}. 
The energy diagram of the eigenstates is shown in Fig.~\ref{fig:lvldiagr}, 
using which we obtain in the second order in $J/U$ \cite{comment}
\bea
H_{\mathrm{eff}} &=& 2 \ve \sum_{j} \hat{m}_j 
+ U \sum_{j} \hat{m}_j \big( 2 \hat{m}_j -1 \big) 
\nonumber \\ & &
+ J^{(2)} \sum_{\expv{j,i}} 
\ck_{j} \, \hat{T}\big( \hat{m}_j, \hat{m}_i \big) \, c_{i} 
\nonumber \\ & &
+ J^{(2)} \sum_{\expv{j,i}} \hat{S}\big( \hat{m}_j, \hat{m}_i \big) , 
\label{HamEff}
\eea
where $J^{(2)} \equiv 2 J^2/U$, and $\hat{T}$ and $\hat{S}$ are defined as
\besa
\hat{T} \big( \hat{m}_j, \hat{m}_i \big) 
&=& \de_{\hat{m}_i \hat{m}_j} \, \sqrt{ \big( 2 \hat{m}_j + 1 \big)  
\big( 2 \hat{m}_i + 1 \big) }  , \qquad \\
\hat{S} \big( \hat{m}_j, \hat{m}_i \big) 
&=& \frac{ \hat{m}_i \big( 2 \hat{m}_j +1 \big) }
{2 \hat{m}_i - \big( 2 \hat{m}_j + 1 \big)} .
\eesa
The third term on the right-hand side of Eq. (\ref{HamEff}) 
describes dimer tunneling between adjacent sites. This tunnel-interaction 
is resonant only between the states of the form 
$\ket{m_j^D} \ket{(m+1)_i^D}$ and
${\ket{(m+1)_j^D} \ket{m_i^D}}$, for which the occupation numbers
of the adjacent sites differ by one; the corresponding matrix
element is equal to $J^{(2)} (m+1) (2 m +1)$. 
The last term of Eq. (\ref{HamEff}), containing the energy shift function 
$\hat{S}$, is responsible for the nearest-neighbor interaction, which, 
depending on the values of $m_j$ and $m_i$, can be positive or negative.  
Adding the two interaction terms between adjacent 
sites $i$ and $j$, we arrive at
\be
\hat{S} \big( \hat{m}_j, \hat{m}_i \big) 
+ \hat{S} \big( \hat{m}_i, \hat{m}_j \big) 
= \frac{2\hat{m}_j^{2} + 2\hat{m}_{i}^{2} + \hat{m}_j + \hat{m}_i}
{4 \big( \hat{m}_j - \hat{m}_i \big)^2 - 1} . \label{attrct-int}
\ee
Thus, when $m_j = m_i$ the interaction between neighboring 
sites is attractive; otherwise it is repulsive. These effects can be
understood as the level shifts of the dimer states, due to 
``level repulsion'' from virtual states having odd occupation numbers. 
The Hamiltonian (\ref{HamEff}) describes the effective dynamics of 
dimers in a 1D, 2D or 3D periodic potential, in the strong 
coupling regime.  Its key features are occupation-dependent
tunneling and nearest-neighbor interactions, as well as strong 
on-site repulsion via the term proportional to $U$. 

It is instructive to consider a 1D configuration 
\[
\cdots \ket{m_{j-2}^D} \ket{m_{j-1}^D} 
\ket{(m+1)_j^D} \ket{(m+1)_{j+1}^D} \cdots ,
\]
which involves the occupation number $m$ of a pair of adjacent 
sites $j-2$ and $j - 1$, and the occupation number $m+1$ 
of sites $j$ and $j+1$. According to Eq. (\ref{attrct-int}), 
there are attractive interactions between sites $j-2$ and $j - 1$,
and between sites $j$ and $j + 1$, while sites $j - 1$ and $j$,
having different occupation numbers, interact repulsively. 
Then the total potential energy is the sum of 
the three terms given by $s = - J^{(2)} \frac{5}{3} (4 m^2 + 6 m + 3)$.
Due to the very large on-site repulsion $\propto U \gg J^{(2)}$, 
the only (near)resonant tunneling interaction of the above state 
is with the state
\[
\cdots \ket{m_{j-2}^D} \ket{(m+1)_{j-1}^D} 
\ket{m_j^D} \ket{(m+1)_{j+1}^D} \cdots ,
\]
having the potential energy $s^{\prime} = J^{(2)} (4 m^2 + 6 m + 3)$.
The corresponding tunneling matrix element $t = J^{(2)} (m+1) (2m + 1)$.
Thus, the ratio of the tunneling (kinetic) energy to the change 
in the potential energy between the above two states is given by 
\be
\frac{t}{s^{\prime} - s} = \frac{3 (m+1) (2m + 1)}
{8 (4 m^2 + 6 m + 3)} .
\ee
This ratio is always smaller than one, its minimal value being 1/8 for $m=0$,
and it quickly approaches a constant $3/16$ for $m > 1$. 

The tunneling $\hat T$ and the nearest neighbor $\hat S$ interactions
are responsible for competing processes: While tunneling favors 
dispersed dimer wavefunctions with long-range coherence, the nearest
neighbor attraction tends to balance the population of neighboring 
sites and to minimize the surface area between regions of different 
occupation number. Since, the interaction term is always larger than 
the competing tunneling term, the ground state will be dominated by 
attractively bound clusters of uniform occupation number and minimal 
surface area, thus representing incompressible ``droplets'' of a 
quantum lattice liquid.

\section{Effective Hamiltonian for $m \leq 1$}

Let us now consider the important special case of a system containing at 
most one dimer per site ($m = 0$ or 1 for all $j$). We thus assume that 
the periodic potential can be loaded initially only with zero or two 
particles per site, at effectively infinite $U/J$ which is then 
adiabatically lowered to a large but finite value, as implemented in 
the optical lattice experiment of Winkler {\it at al.} \cite{KWEtALPZ}. 
Just as dimers are energetically forbidden to dissociate in the absence 
of dissipation, the single-site dimer occupation numbers will never 
exceed unity, for this would require a large energy input of the order 
of $5 U$. Under these conditions, the effective Hamiltonian (\ref{HubHam})
can be recast simply as
\bea
H_{\mathrm{eff}}^{(0,1)} 
&=& \big[2 \ve + U + 2 d J^{(2)} \big] \sum_j \hat{m}_j
+ J^{(2)} \sum_{\expv{j,i}} \ck_{j} c_{i} 
\nonumber \\ & &
- 4 J^{(2)} \sum_{\expv{j,i}} \hat{m}_j \hat{m}_i , 
\label{HamEff01}
\eea
where the only allowed values of $m$ are 0 or 1. Thus, in addition to the 
tunneling interaction with negative effective mass, there is a stronger 
attractive interaction between dimers localized at neighboring sites, 
which can bind them together as discussed in Appendix \ref{app:2dmrwv}. 
Note that (\ref{HamEff01}) has the form of an extended Hubbard model, 
like that which describes electrons in a crystal lattice or quantum 
dot array \cite{GNDPPL}. There, however, the nearest-neighbor 
interaction is repulsive, while in our case it is attractive. 
We also note that related effects have been predicted for strongly 
interacting mixtures of bosonic and fermionic atoms in an optical 
lattice \cite{LSBF}, wherein the fermions tend to pair with one or 
more bosons, forming composite fermions with nearest-neighbor interaction. 

\begin{figure}[t]
\centerline{\includegraphics[width=8.5cm]{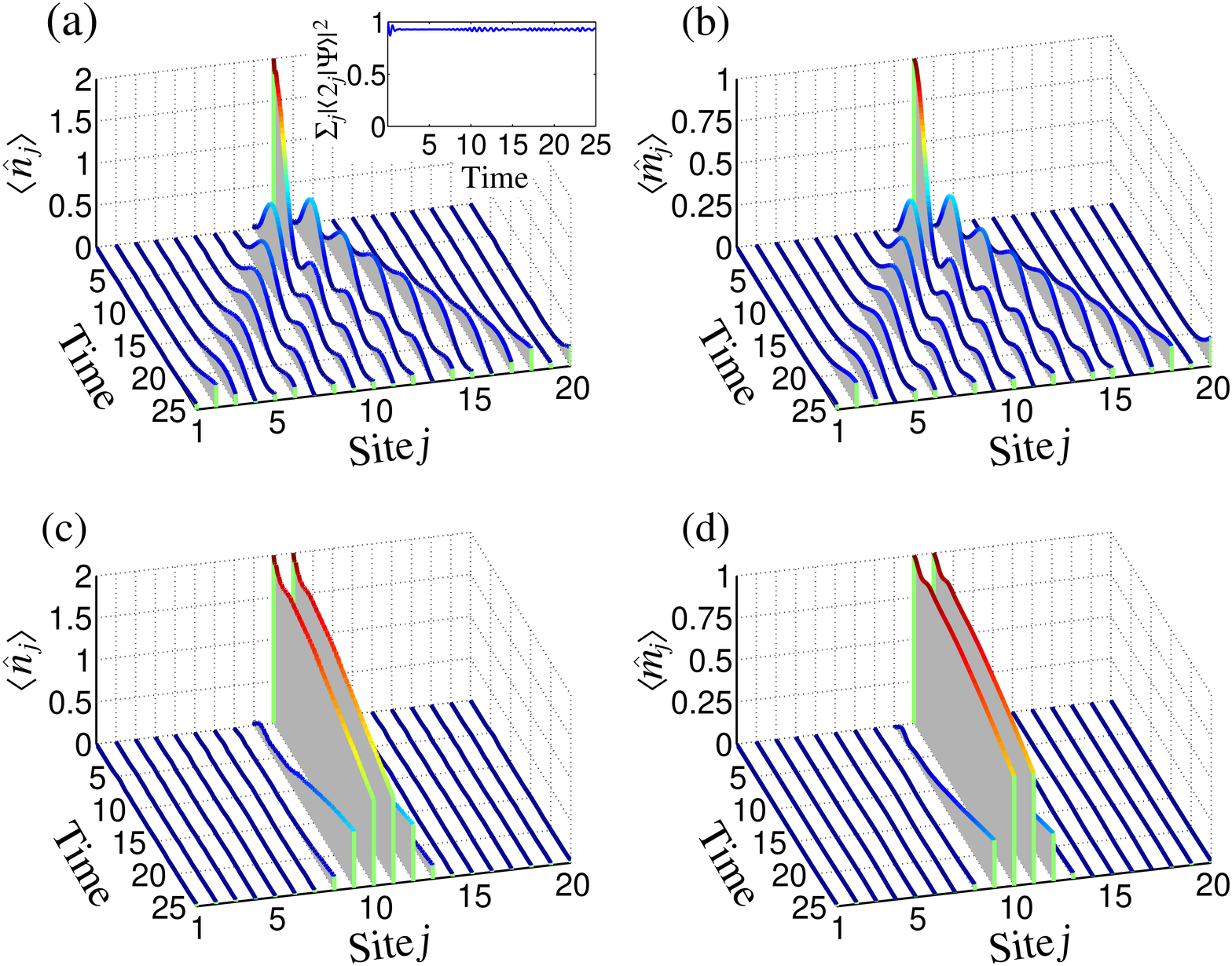}}
\caption{Dynamics of one dimer, (a) and (b), and two dimers, (c) and (d), 
in a 1D lattice of 20 sites, for $J/U = 0.1$.
(a) and (c) are numerical solutions of the Schr\"odinger
equation with the Bose-Hubbard Hamiltonian (\ref{HubHam}),
while (b) and (d) are obtained with the effective Hamiltonian
(\ref{HamEff}) [or (\ref{HamEff01})]. Inset in (a) shows
the time-evolution of $\sum_j |\braket{2_j}{\Psi}|^2$, where
$\ket{\Psi(t)}$ is the system wavefunction.
Time is in units of $J^{-1}$.}
\label{fig:12dimdyn}
\end{figure}

To verify the validity of our perturbative approach
in the limit of $J/U \ll 1$, we have numerically solved the 
Schr\"odinger equation for the cases of one and two dimers 
in a 1D lattice of 20 sites, using the Bose-Hubbard Hamiltonian 
(\ref{HubHam}), and the effective Hamiltonian (\ref{HamEff}) 
[or (\ref{HamEff01})]. As shown in Fig.~\ref{fig:12dimdyn}, 
the dynamics of the system obtained from the exact and effective 
Hamiltonians is very similar; the difference between the exact 
and effective models decreases for smaller values of $J/U$, as expected.
In the inset of Fig.~\ref{fig:12dimdyn}(a) we plot the projection of 
the system wavefunction $\ket{\Psi(t)}$ onto the states $\ket{2_j}$ 
with two particles per site. As seen, $\sum_j |\braket{2_j}{\Psi}|^2 \simeq 1$
at all times, attesting to the fact that the two particles forming a dimer 
are strongly bound to each other, even though the center-of-mass wavefunction 
of the dimer disperses with time due to the tunneling $J^{(2)}$. 
Figs.~\ref{fig:12dimdyn}(c),(d) reveal the greatly reduced dispersion 
for a pair of neighboring dimers attractively bound to each other 
(see Appendix \ref{app:2dmrwv}): the two-dimer pair collectively 
tunnels resonantly only at fourth order in the fundamental $J$ 
(second order in $J^{(2)}$).

\begin{figure}[t]
\centerline{\includegraphics[width=6cm]{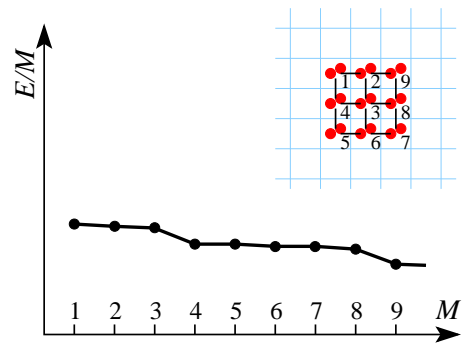}}
\caption{Energy per dimer $E/M$ versus the number of dimers $M$ forming 
a cluster in 2D square lattice. As seen, $E/M$ abruptly drops once a 
square droplet with the dimension $\sqrt{M} \times \sqrt{M}$ is formed,
since the addition of the last dimer results in the formation of two 
``bonds''.}
\label{fig:2Dclast}
\end{figure}

The above reasoning can be extended to the case of more dimers. Since 
each dimer is attracted to its immediate neighbor, for a given number 
of dimers, the configuration that minimizes the energy of the system 
would correspond to clustering of the dimers together in such a way as 
to maximize the number of the nearest--neighbor (attractive) interactions. 
Thus, in 1D all the dimers would stick together in a line without voids, 
while for 2D or 3D square lattice, the dimers would tend to arrange 
themselves in a square (2D) or a cube (3D), as shown in Fig.~\ref{fig:2Dclast}.
(Because of the discretized perimeter metric in the lattice, minimal
surfaces of these ``droplets'' are rectangular rather than round, 
however large.)

\section{Phase diagram of the grand canonical ensemble}

\begin{figure}[b]
\centerline{\includegraphics[width=8cm]{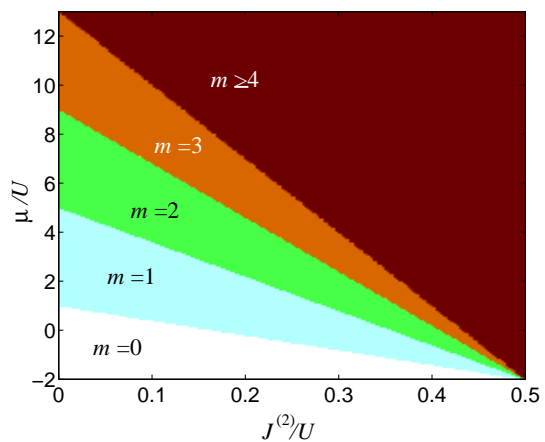}}
\caption{Phase diagram of the grand canonical ensemble 
obtained from exact diagonalization of Eq. (\ref{GCHam}). 
The Hilbert space is restricted by five sites (periodic boundary conditions),
with each site occupation number in the range of $0 \leq m \leq 4$.
The areas of integer filling are tightly adjoined to each other, 
with no significant extent of fractional filling phase.}
\label{fig:phsdgrm}
\end{figure}

In order to understand the ground-state properties of the effective
Hamiltonian (\ref{HamEff}), we consider the grand canonical ensemble 
described by the operator
\be
K = H_{\mathrm{eff}} - \mu \sum_j \hat{m}_j , \label{GCHam}
\ee
where $\mu$ is the chemical potential assumed uniform for all sites.
The corresponding phase diagram, calculated numerically for a small
1D lattice at zero temperature, is shown in Fig.~\ref{fig:phsdgrm}. 
Since the tunneling interaction is always smaller than the attractive 
interaction between neighboring sites with equal occupation numbers, 
we observe only incompressible phases, with uniform, commensurate filling. 
All systems with incommensurate dimer filling lie on the border lines 
between the incompressible phases, which verifies the qualitative 
discussion above. When adding a dimer to the system, it is energetically
favorable for this dimer to be bound to an already existing cluster 
or droplet rather than to move freely. 

This picture changes, however, when a finite temperature $T$ is
considered. If $T$ is sufficiently large the minimum free energy
may be attained when the dimers move freely rather than being bound to a
cluster. We thus expect the system to show a first-order phase 
transition from a ``quantum-droplet'' phase  to a ``gas'' phase
at some critical temperature $T_c$. 

\section{Droplet--gas transition}

\begin{figure}[t]
\centerline{\includegraphics[width=8cm]{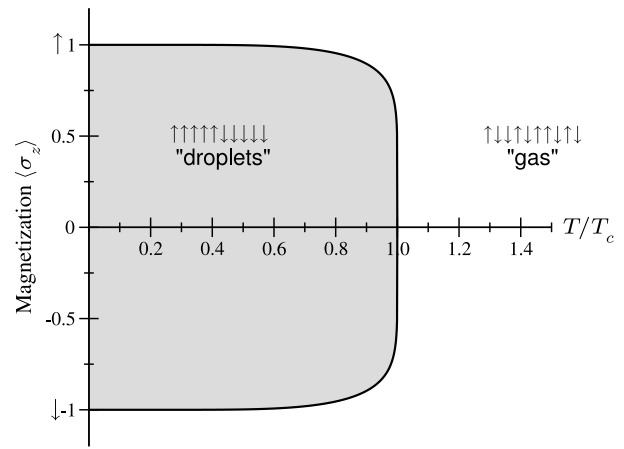}}
\caption{Temperature phase diagram of the 2D Ising model. In the shaded
area, the ferromagnetic spin domains are formed. As temperature is 
increased, for $\expv{\sigma^z} \neq 0$, the system undergoes a first-order
phase transition to the ``gas'' phase, while at $\expv{\sigma^z} = 0$ 
it is a second order phase transition. See text for more details}
\label{fig:2DIsing}
\end{figure}

The system described by the effective Hamiltonian $H_{\mathrm{eff}}^{(0,1)}$
is equivalent to the well-known spin-$\frac{1}{2}$ $XXZ$ model in a 
magnetic field \cite{Sadchev,Takahashi}. Indeed, with the mapping 
$\ket{0_j} \to \ket{\ds_j}$ and $\ket{1_j} \to \ket{\us_j}$ and 
simple algebraic manipulations, which essentially amount to the  
Wigner-Jordan transformation, Eq.~(\ref{HamEff01}) can be cast as
\bea
H_{\mathrm{spin}}  
&=& C - 2 h_z \sum_j \sigma_j^z + \frac{J^{(2)}}{4} \sum_{\expv{j,i}} 
\big( \sigma_{j}^x \sigma_{i}^x + \sigma_{j}^y \sigma_{i}^y \big) 
\nonumber \\ & &
- J^{(2)} \sum_{\expv{j,i}} \sigma_{j}^z \sigma_{i}^z , 
\label{HamSpin}
\eea
where $C$ is an immaterial constant,  
$h_z = 2 d J^{(2)} - \frac{1}{4} \big[2 \ve + U + 2d J^{(2)} \big]$
is an effective magnetic field, and $\sigma_{j}^x$, $\sigma_{j}^y$ 
and $\sigma_{j}^z$ are the Pauli spin matrices. Note that, unlike the
usual situation in spin systems, here the averaged ``magnetization'' 
of the system $\expv{\sigma^z}$ is fixed by the condition 
$\expv{m} = (1 +\expv{\sigma^z})/2$, where $\expv{m}$ is the dimer 
filling factor. In this description, we have a ferromagnetic spin coupling
described by the last term of Eq.~(\ref{HamSpin}), which dominates over the
spin-exchange interaction. At low temperatures ($k_{\textrm{B}} T < J^{(2)}$),
the ``spins'' therefore form a ferromagnetic domain with the spins pointing 
up, surrounded by the remaining spins pointing down. At certain critical
temperature $T_c$, the spin domains disappear and a random distribution of
the $\ket{\us_j}$ and $\ket{\ds_j}$ states emerge. In order to estimate
$T_c$, we note that in the above spin Hamiltonian the $ZZ$ coupling is 
significantly larger than the $XX$ and $YY$ couplings, which, to a reasonable
approximation, can be neglected. Equation~(\ref{HamSpin}) then reduces to 
the Ising Hamiltonian \cite{IsingBook}, whose analytic properties in 2D 
are well known. In Fig.~\ref{fig:2DIsing} we show the finite--temperature 
phase diagram of the 2D Ising model. The shaded ferromagnetic spin domains 
at low temperatures correspond to the ``droplets'' of our model. The boundary
of that region $\expv{\sigma^z}_c (T)$ is defined through
\[
\expv{\sigma^z}_c (T)= \left[ 1 - 
\sinh^{-4} \left( \frac{2J^{(2)}}{k_{\textrm{B}} T} \right) \right]^{1/8} .
\] 
As temperature is increased, for $\expv{\sigma^z} \neq 0$ the system 
undergoes a first-order phase transition from the ``droplet'' to the ``gas''
phase. For $\expv{\sigma^z} = 0$, the transition is a monotonous second 
order phase transition, for which the critical temperature $T_c$ corresponds 
to $\expv{\sigma^z}_c (T_c) =0$ which yields 
$k_{\textrm{B}} T_c/J^{(2)}= 2/\textrm{arcsinh}(1)=2.2692$

\section{Conclusions and closing remarks}

Before closing, let us briefly consider several important 
experimental issues. As we have stated in the beginning of this
paper, the most relevant experimental situation for the present 
study is realized by cold bosonic atoms loaded into an 
optical lattice \cite{KWEtALPZ}. Initially, pairs of atoms 
($^{87}$Rb) are adiabatic converted with near unit efficiency into 
chemically bound molecules (Rb$_2$) using magnetic field sweep 
across a Feshbach resonance. This step is then followed by 
removing all chemically unbound atoms with the combined 
radio-frequency and optical purification pulses. Finally, 
the dimer molecules are adiabatically converted back into pairs 
of atoms localized at the same site, with no significant
admixture of unpaired atoms. In the case of strong on-site 
repulsion $U \gg J$, these pairs of atoms form the dimers 
studied in this paper. When the lattice sites
are occupied by more than one dimer, the three- and four
body collisions will presumably be the dominant loss mechanism 
for the atoms. In a recent study, Campbell {\it et al.} \cite{n12345}
have experimentally realized a Mott insulator phase of cold $^{87}$Rb 
atoms with per site particle numbers of $n =1,2,3,4,5$ in successive 
spatial shells, and determined the lifetime of each shell. 
The observation for $n=2$ was around 100~s, and for $n \geq 3$  
around $0.5$~s. On the other hand, the rate of dimer tunneling 
$J^{(2)}$ estimated from \cite{KWEtALPZ} is about $10-20$~s$^{-1}$ 
which is thus three orders of magnitude larger than the loss rate 
for $n = 2$ (i.e., $m = 1$), and an order of magnitude larger than 
the loss rate for $n = 4$ (i.e., $m = 2$).

In the experiment of Winkler {\it et al.} \cite{KWEtALPZ}, in order to 
determine the fraction of the remaining dimers for various experimental 
conditions and hold times, the authors repeat the above sequence (i.e., 
conversion of atoms pairs into molecules, purification, and reverse conversion)
and then use the conventional absorption imaging. With minor modification,
this method can be employed to experimentally verify the formation
of clusters of dimers. Recall that dimers forming a cluster become
immobile, while individual unbound dimers are mobile, moving around
the lattice with the tunneling rate $J^{(2)}$. Let us assume that at 
the boundaries of the lattice of linear dimension $l$ there exists 
some dimer loss mechanism (see below). Then, if the dimers are not 
bound to each other, after a sufficient time of the order of 
$t_{\textrm{escape}} \sim l/J^{(2)}$, they will escape from the lattice, 
while immobile dimers bound in a cluster will remain in the lattice,
which can be verified by the same absorption imaging. The loss mechanism
at the boundaries of the lattice can be an atom evaporation by focused 
laser beams. Alternatively, if the lattice potential is created by strongly 
focused (blue-detuned) laser field, then away from the central region, 
as the intensity of the field falls off, the tunneling barriers become 
lower. As a result, the dimer mobility increases, and eventually even 
individual atoms can move practically freely, quickly escaping the lattice. 

Let us finally note that in the above discussion on the properties 
of repulsively bound pairs of particles in a periodic potential, 
we have neglected the effects of energy dissipation in the system. 
Assuming small temperature and the dimer filling factor 
$\expv{m} \leq 1/2$ (average particle filling factor $\expv{n} \leq 1$), 
it is obvious that in the presence of energy relaxation with 
a characteristic rate $\ga$ (such as from spontaneous emission of phonons 
in a solid, or inelastic collisions with a cold background gas for 
atoms in an optical lattice), the lifetime of repulsively bound pairs 
will be limited by $\ga^{-1}$. But for an initial random distribution 
of dimers in the lattice, dissipation on shorter time scales than 
$\ga^{-1}$ will drive formation of multi-dimer clusters, to minimize the 
energy of the dimer system.  Furthermore, once a cluster is formed, 
dimer dissociation becomes a surface process only, because dissociation
of a dimer inside the cluster would mean forming a ``trimer'' at an adjacent 
site, which requires energy input $U$, instead of energy release.  
Note also that the collision of a single unpaired particle with a dimer 
involves resonant single-particle exchange \cite{comment}. The admixture 
of single particles thus brings a complicated interplay between dimer 
dissociation and bound dimer collisions with single particles. 
Detailed understanding of fluctuations and dissipation in the liquid-like 
phase of clustered dimers will require further investigation, bringing
the physics of first-order phase transitions into the arena of ultracold atoms.

\begin{acknowledgments}
This work was supported by the EC RTN EMALI. D.~P. acknowledges
financial support from the Alexander von Humboldt Foundation
during his stay at the TU Kaiserslautern, where most of this
work was performed. 
\end{acknowledgments}

\appendix

\section{Perturbative derivation of a dimer wavefunction}
\label{app:1dmrwv}

The exact wave function and dispersion relation for single dimers can 
be obtained analytically in 1D \cite{KWEtALPZ,molmer}. As a tutorial 
for our derivation of the effective many-dimer Hamiltonian, we analyze
the single dimer  perturbatively for small $J/U$ \cite{bosMI}. Given a 
dimer centered at site $j$, in 1D its ``internal'' state $\ket{D_j}$ is
\be
\ket{D_j} = A_{j,0} \ket{2_j} 
+ \sum_{r} \big( A_{j,r} \ket{1_j} \ket{1_{j+r}} 
+ A_{j,-r} \ket{1_{j-r}} \ket{1_j} \big), \label{DmrWFnc}
\ee 
where $r= 1,2,\ldots$ is the distance in sites one of the 
constituent particles of the dimer has tunneled away from the other. 
In zeroth order in $J$, we have $A_{j,0} = 1$ and all $A_{j, \pm r} =0$. 
At in the successive higher orders in $J/U$ it is easy to see that 
$A_{j, \pm r} \simeq  \sqrt{2} \left( -\frac{J}{U} \right)^{r} A_{j,0}$.
The corresponding probability of finding the dimer constituents separated by
$r$ sites is 
$P_{j,r} = |A_{j,r}|^2 + |A_{j,-r}|^2 
= 4  P_{j,0} \left( \frac{J^2}{U^2} \right)^{r}$,
while $P_{j,0} = |A_{j,0}|^2$. For $|J/U| \ll 1$, the normalization 
condition $\sum P_{j,r} = 1$ then yields
\be
A_{j,0} \simeq \sqrt{\frac{U^2 - J^2}{U^2 + 3 J^2}} , \quad 
A_{j, \pm r} \simeq  (-1)^r \sqrt{2} A_{j,0}
\left(\frac{J}{U} \right)^{r} .
\ee
Note the alternating sign of the amplitudes $A_{j, \pm r}$ between the 
sites $r$. Expressing the tunneling probabilities $P_{j,r}$ as
\be
P_{j,r} = 
4 P_{j,0} \, \exp \left[\ln \left(\frac{J^2}{U^2} \right)^{r} \right] 
= 4 P_{j,0} \, e^{-r/\zeta} , 
\ee 
we find the localization (or ``bond'') length of the dimer to be 
$\zeta = [2\ln (U/J)]^{-1}$, so that $\zeta<1$ for $U / J > \sqrt{e}$.
These results agree with the exact expressions \cite{KWEtALPZ,molmer} 
in the limit $J \ll U$, and they can be extended to higher dimensions, 
which are less tractable by the exact methods. Thus, for example, 
in 2D we obtain 
\bea
A_{j,0} &\simeq & \sqrt{\frac{U^2 - 3 J^2}{U^2 + 5 J^2}} , \\
P_{j,r} &\simeq & 8 P_{j,0} 
\left[ \frac{\Gamma(r+\frac{1}{2})}{\sqrt{\pi} \Gamma(r+1)} 4^r -1 \right] 
\left(\frac{J^2}{U^2} \right)^{r} 
\nonumber \\ & & 
< 8 P_{j,0} \, \exp \left[\ln \left(\frac{4 J^2}{U^2} \right)^{r} \right] 
= 8 P_{j,0} \, e^{-r/\zeta} , \quad
\eea
where the localization length is $\zeta = [2\ln (U/2J)]^{-1}$.

\section{Perturbative derivation of two-dimer wavefunction}
\label{app:2dmrwv}

Consider a 1D configuration with two dimers occupying adjacent sites 
$\expv{j,i}$. Their potential energy is lower by the amount $8 J^{(2)}$ 
than that of two dimers separated by one or more lattice sites 
[see Eq. (\ref{HamEff01})]. In analogy with the case of two particles
forming a dimer, we can calculate the wavefunction $\ket{Q_{ji}}$ of the 
attractively bound dimer pair perturbatively in the effective tunneling 
$J^{(2)}$. To that end, we expand the wavefunction $\ket{Q_{ji}}$ as
\bea
\ket{Q_{ji}} &=& B_{ji,0} \ket{1^D_j} \ket{1^D_i} 
+ \sum_r \big( B_{ji,r} \ket{1^D_j} \ket{1^D_{i+r}} \nonumber \\ & & \qquad
+ B_{ji,-r} \ket{1^D_{j-r}} \ket{1^D_{i}} \big),   
\eea
where $r = 1,2,\ldots$ is the number of sites separating the dimers.
We then obtain $B_{ji,\pm r} \simeq (-1/8)^r B_{ji,0}$, which, upon
requiring the normalization $\sum P_{ji,r} = 1$, where 
$P_{ji,r} = |B_{ji,r}|^2 + |B_{ji,-r}|^2$, yields
\be
B_{ji,r} \simeq \sqrt{ \frac{63}{65}} \left(- \frac{1}{8} \right)^r .
\ee
We therefore have $P_{ji,r} \simeq 2 e^{-r/\xi}$ with the localization 
length $\xi = (\ln 64)^{-1} \simeq 0.24$. Hence, two dimers localized 
at adjacent lattice sites are closely bound to each other. It can be
shown that this conclusion also holds in 2D and 3D.


\begin{thebibliography}{99}
\bibitem{comment} Note that states of the form 
${\ket{n_j}\ket{ (n \pm 1)_i}}$ and ${\ket{(n \pm 1)_j}\ket{n_i}}$
 (with $\expv{j,i}$) are resonantly coupled to each other 
via single particle tunneling. But by assuming that only even 
number of particles per site are present initially
in the system, we have explicitly excluded such events.
\end{thebibliography}
\end{document}